\documentclass[preprint,showkeys,aps,prb]{revtex4}   
\usepackage[english]{babel}
\usepackage[dvips]{graphicx}  
\usepackage{color}                         
\begin{document}


\title{Ergodic Time-Reversible Chaos for Gibbs' Canonical Oscillator}

\author{
{\bf William Graham Hoover}, Ruby Valley Research Institute          \\
Corresponding Author email : hooverwilliam@yahoo.com                 \\
Ruby Valley Research Institute                                       \\
Highway Contract 60, Box 601, Ruby Valley, Nevada 89833, USA ;       \\
{\bf Julien Clinton Sprott}, Department of Physics                   \\
University of Wisconsin - Madison, Wisconsin 53706, USA  ;           \\
{\bf Puneet Kumar Patra}, Advanced Technology Development Center     \\
Department of Civil Engineering, Indian Institute of Technology      \\
Kharagpur, West Bengal, India 721302 .                               \\
}
\date{\today}

\keywords{Ergodicity, Chaos, Algorithms, Dynamical Systems}


\vspace{0.1cm}

\begin{abstract}
Nos\'e's pioneering 1984 work inspired a variety of time-reversible deterministic
thermostats.  Though several groups have developed sucessful {\it doubly}-thermostated
models, single-thermostat models have failed to generate Gibbs' canonical
distribution for the one-dimensional harmonic oscillator.  A 2001 doubly-thermostated
model, claimed to be ergodic, has a singly-thermostated version.  Though neither of
these models {\it is} ergodic this work has suggested a successful route toward
singly-thermostated ergodicity.  We illustrate both ergodicity and its lack for these
models using phase-space cross sections and Lyapunov instability as diagnostic tools.

\end{abstract}

\maketitle

\section{Single-Variable Thermostats and Gaussian Ergodicity}

In 1984 Hoover explored the application of the Nos\'e-Hoover version\cite{b1} of Nos\'e's
canonical motion equations\cite{b2,b3} to a harmonic oscillator at thermal equilibrium
with coordinate $q$ , momentum $p$ , temperature $T$ , and thermostat variable $\zeta$ :
$$
\{ \ \dot q =p \ ; \ \dot p = -q -\zeta p \ ; \
\dot \zeta = [ \ p^2 - T \ ]/\tau^2 \ \} \ {\rm [ \ NH \ ]} \ . 
$$
Posch, Hoover, and Vesely found that this model partitions the $(q,p,\zeta)$ phase space
into many separate toroidal regions embedded in a chaotic sea\cite{b4}. The complexity
and the stiffness of the solutions increase rapidly as the thermostat response time
$\tau$ is reduced.  In addition to equilibrium applications analogous motion equations
can be used to thermostat irreversible nonequilibrium simulations such as steady shear
and heat flows.  The harmonic oscillator can generate steady-state heat flow
problems if the temperature varies in space\cite{b5,b6} :
$$
1 - \epsilon < T = T(q) = 1 + \epsilon\tanh(q) < 1 + \epsilon \ .
$$
\textcolor{black}{Here $\epsilon$ is the maximum value of the temperature gradient, $(dT/dq)$,
to which the oscillator is exposed. It can be viewed as the strength of nonlinearity, and
depending on its value, one can move from the equilibrium regime (where $\epsilon = 0$) to
the nonequilibrium regime (where $\epsilon > 0$).}

Somewhat paradoxically, the Nos\'e-Hoover  motion equations as well as all the others we
consider here  are {\it time-reversible}, even away from equilibrium.  That is, any time-ordered
sequence of $(q,p,\zeta)$ points can be reversed either by [ 1 ] changing the sign of $dt$
in the integrator, or [ 2 ] by changing the signs of the $(p,\zeta)$ variables.
The harmonic oscillator equations also have mirror symmetry.  Changing the signs
$(+q,+p) \longleftrightarrow (-q,-p)$ gives an additional pairing of solutions.

\textcolor{black}{Apart from being time-reversible, a good thermostat must result in \textit{ergodic} dynamics. Ergodicity of the dynamics connects dynamical averages with corresponding Boltzmann-Gibbs phase averages. In describing the results of the present work, we have used Ehrenfests' idea of ``quasiergodicity'', where the dynamics eventually comes arbitrarily close to each feasible point, interchangeably with ``ergodicity''.}

For the equilibrium \textcolor{black}{Nos\'e-Hoover harmonic oscillator, the Gaussian distribution is the stationary solution of the Liouville's phase-space continuity equation:}
$$
v = \dot r = (\dot q,\dot p,\dot \zeta) \longrightarrow
(\partial f/\partial t) = -\nabla_r\cdot (fv) \equiv 0 \longrightarrow
$$
$$
f(q,p,\zeta) \propto e^{-q^2/2T}e^{-p^2/2T}e^{-\zeta^2\tau^2/2T} \ ,
$$
On the other hand, numerical work gives two kinds of solutions, either quasi-periodic tori, \textcolor{black}{with all Lyapunov exponents being zero} or
a single  chaotic, Lyapunov-unstable sea. \textcolor{black}{The global dynamics, therefore, either remains confined within the tori, or occupies the chaotic sea separated from the tori, depending upon the initial conditions,
the temperature $T$ , and the response time $\tau$. In other words the presence of two sets of global maximal Lyapunov exponent - one positive and another zero, indicates that a trajectory starting from an arbitrary initial condition is unable to explore the neighbourhood of the entire feasible phase-space. As a result, the phase-space gets partitioned into at least two noncommunicating regions, violating the metric indecomposibility of the phase space -- the necessary and sufficient condition for ergodic dynamics according to Birkhoff's theorem.} Thus the
singly-thermostated oscillator equations are not ``ergodic'', so that Gibbs' statistical
mechanics is unable to describe the oscillator's properties.  For the next 15 years, which
included many failed attempts, no singly-thermostated oscillator models were found to be
ergodic.

This letter announces our recent achievements toward the longstanding goal of ergodic
singly-thermostated oscillator models.  We have carried
out a comprehensive exploration of a previous model claimed to be ergodic, and found that
it is not.  As a result of those investigations we have found a path leading to a
singly-thermostated and physically motivated  ergodic model for the harmonic oscillator.
We lay out the details of these discoveries in what follows and encourage the reader to
help explore the new areas opened up by our work. 

\section{Ergodicity is Typically Absent in the SF Model}

In 2001 Sergi and Ferrario [ SF ] announced that they had found an ergodic
thermostated  oscillator model\cite{b7}. \textcolor{black}{In addition to the oscillator coordinate, momentum, and thermostat variable $(q,p,\zeta)$ their model includes a parameter $\nu$ which can be either positive or negative:
$$
\dot{q}=p(1+\zeta \nu) \ ; \ \dot{p} = -q - \zeta p \ ; \ \dot{\zeta} = [ \ p^2 - T - qp\nu \ ]/\tau^2 \ ;
\  \dot{\eta} = \zeta\ .
\nonumber
\label{eq:SF_originial}
$$
}Here, and in what follows, we will ignore the
fact that SF actually solve the above {\it four} equations, not just the three shown below:  
$$                                         
\{ \ \dot q = p(1 + \zeta \nu) \ ; \ \dot p = -q -\zeta p \ ; \
\dot \zeta = p^2 - T - qp\nu \ \} \ {\rm [ \ SF \ ]} \ ,
$$
This is because their work was based on a Hamiltonian with two degrees of freedom. \textcolor{black}{Consider a particular initial condition ($q,p,\zeta$) that evolves in some time $t$ to a unique ($q^\prime,p^\prime,\zeta^\prime$). The latter variables do not depend on the initial value of $\eta$, which could be given or not, arbitrarily.} The fourth equation, for the evolution of a variable which is the time integral of $\zeta$ , plays no role at all in the dynamics of $(qp\zeta)$ and can so be ignored, which we do throughout. This extraneous variable obscured the fact that SF implicitly claimed ergodicity for a {\it singly}-thermostated
oscillator. As a result, this desirable feature of their relatively widely-cited paper has been previously ignored. However, as a consequence of removing $\dot{\eta}$, the symplecticity of the dynamics disappears.

Like the NH model the SF oscillator has mirror symmetry
$
(+q,+p) \longleftrightarrow (-q,-p)
$ .  In addition the time reversibility of the Sergi-Ferrario equations requires that the
functions $p$ and $\zeta$ , {\it as well as the parameter} $\nu$ , all change sign in
the reversed motion with the coordinate values unchanged. For clarity we have replaced Sergi
and Ferrario's parameter ``$\tau$'' by $\nu$ throughout the present work.  This change emphasizes
that an increase in $| \ \nu \ |$ reduces the response time of the thermostat terms.

\textcolor{black}{For the remainder of this study, we choose to keep $\tau = 1$. Usually $\tau$ , which represents the relaxation time of the dynamics, is chosen according to the relation\cite{b12}: $\tau^2=kT/\omega^2$, where $\omega$ is the angular frequency of the system. In our present case, since the system comprises a single harmonic oscillator with unit mass and spring constant, $\omega = 1$. Additionally, most of the work ascertaining the ergodicity of thermostatted dynamics has taken the relaxation time to be unity. We wish to highlight the fact that if the relaxation time is chosen too large, it will have no effect on the system dynamics, while if $\tau$ is chosen too small, the equations become too stiff.}

Sergi and Ferrario claimed that their four [ but actually only three, for the reason just cited ]
oscillator equations\cite{b7} were ergodic ( filling
out the entire three-dimensional Gaussian distribution ) for  $\nu > 0.5$ . That
surprising claim sparked the present work. To begin our exploration of their model we
carried out a simulation of the SF equations with the temperature $T$ and parameter $\nu$
both equal to unity and with the initial conditions $(q,p,\zeta) = (1,1,1)$ . {\bf Figure 1}
shows the resulting torus, colored according to the local flow instability.  Evidently
this special case of the SF model is definitely {\it not} ergodic.

\textcolor{black}{The difficulty in isolating a small embedded torus by looking at the global dynamics \cite{b13} prompted us to investigate the Poincar\'e section at $\zeta = 0$. In fact, any other \textit{typical} Poincar\'e section would have served our purpose. Recall that Gibbs' probability density is Gaussian in both $q$ and $p$. Accordingly sections in $q$ and $p$ (as well as in $\zeta$) that are far from origin are atypical, and may not give any useful results. So long as the section chosen is a typical one, the dynamics within it can be studied to understand ergodicity.}
 
\begin{figure}
\vspace{1 cm}
\includegraphics[width=4.0in]{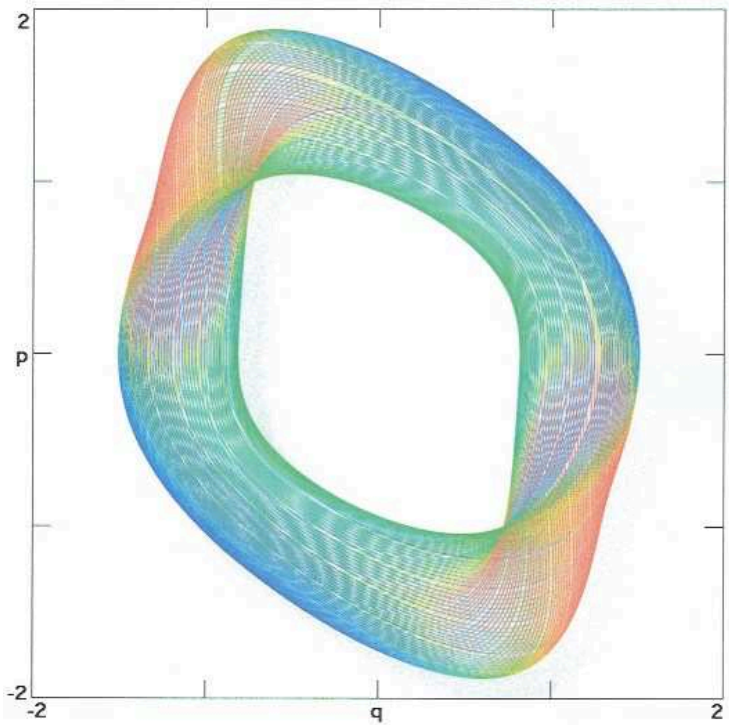}
\caption{
The torus shown here results from the intial conditions $(q,p,\zeta) = (1,1,1)$ using
Sergi and Ferrario's original equations with $\nu = +1$ .  The local values of the
largest Lyapunov exponent on the torus are indicated by color :
$-1.06 \ {\rm \ (blue) \ } < \lambda_1(t) < 1.89 \ {\rm \ (red) \ }$ .  Its time-averaged
mean value, $\lambda_1 = \langle \ \lambda_1(t) \ \rangle $ is zero.  The temperature is unity.
}
\end{figure}

Rather than abandoning the SF approach we also looked for modifications that might
be ergodic.  Changing the parameter $\nu $ from 1 to 2 or 3 or 4 or 5 or 6 and applying
due diligence led in each case to the discovery of nested tori.  Typically the tori
penetrate the plane $\zeta = 0$ in four widely-separated distinct places.  {\bf Figure 2}
illustrates these ``period-four'' equilibrium points for the SF equations.  Just as
in the other Figures the online version is colored according to the local value of
the largest of the three Lyapunov exponents, $\lambda_1(t)$.  We denote the long-time
average value of this exponent by $\lambda_1 \equiv \langle \ \lambda_1(t) \ \rangle $ .

Holes in the chaotic sea are most easily found visually.  Then, zooming in on such a hole
the central point corresponding to a periodic orbit can be found.  By first looking at cross
sections decorated by a million penetration points and then zooming in on the holes we can
obtain precise six-figure estimates for the  $(q,p,0)$ points that lie at the center of
each hole, on the central periodic orbit.  Viewed in the $(q,p,0)$ plane, diligent searches
showed that the six choices of $\nu$ shown in Figure 2 , {\it all} include simple tori
centered on a periodic \textcolor{black}{orbit} and embedded in a chaotic sea.  Looking at the Figure, the
relatively small but clearly visible holes can be seen for $\nu = 2,4,5,$ and $6$ . The
large irregular holes for $\nu = 1$ form a cross section of the torus shown in Figure 1.
The four tiny holes corresponding to $\nu = 3$ are too small to see without zooming in.  Some
of the details of these investigations are described in what follows, along with a
concluding Summary, Discovery, and Advice section.

\begin{figure}
\vspace{1 cm}
\includegraphics[width=4.0in]{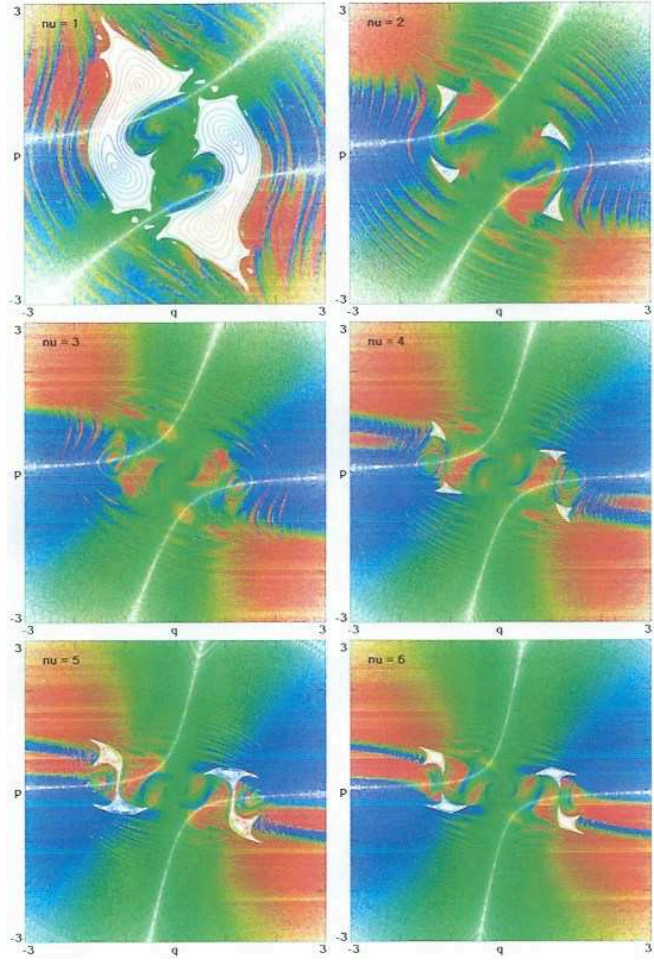}
\caption{
\textcolor{black}{Cross sections at $\zeta = 0$, for different values of $\nu$, with the local values of the largest Lyapunov exponent $\lambda_1(t)$ colored from $\simeq - 1.0$ (blue) to $\simeq + 1.0$ (red). The top, middle and bottom figures on the left correspond to $\nu = 1, 3$ and 5, respectively. Likewise, the figures on the right correspond to $nu = 2,4$ and 6, respectively.} The white curves correspond to intersections with the nullcline
surface.  There the phase-point velocity is parallel to the $(q,p,0)$ plane with $q = (p^2 - 1 )/(\nu p)$ . The tori for $\nu = 2,4,5$ and $6$ penetrate the plane within four roughly-triangular holes in those cross sections. The tori for $\nu = +3$ are much
smaller, as is detailed in Figure 3. The four penetrations occur as two pairs of mirror-image points : $(q,p,0) = (\pm 1.04031,\pm 0.39432,0)$ and $(\pm 1.18488,\mp 0.94527,0)$ .  The temperature is unity throughout.
}
\end{figure}

\section{Lyapunov Instability and Gaussian Moments}

The largest time-averaged Lyapunov exponent $\lambda_1$ measures the exponential tendency
for two nearby chaotic trajectories to separate,
$\delta (t) \simeq \delta(0)e^{+\lambda t}$ . The local value $\lambda_1(t)$ exhibits
fluctuations, even in the regular toroidal regions where the long-time average,
$\lambda_1$ , is zero.  $\lambda_1$ is positive in the
chaotic sea.  It is a measure of the chaos there.  For online viewing we have included the
local values with color, ranging from blue to
red as the exponent increases.  For the model of {\bf Figure 1} the long-time-averaged
exponent is equal to zero, as expected for a two-dimensional torus in a three-dimensional
space.  Simulations with $\nu = 2,3,4,5,6$ looked much more promising, as they all generated
``fuzzy balls'' in $(q,p,\zeta)$ space. \textcolor{black}{We picture the three-dimensional Gaussian distribution, proportional to $e^{-q^2/2-p^2/2-\zeta^2/2}$, as a fuzzy ball. It is evident that the density falls off exponentially in all directions as one moves away from the ``center'' of the pictured ball.}

We next investigated the ergodicity of these fuzzy balls by measuring the time-averaged
moments $\{ \ \langle \ q^2,q^4,q^6,p^2,p^4,p^6 \ \rangle \ \}$ . For every value of
$\nu$ , using a spacing of 0.05 with $0 < \nu \leq 6$ , we found \textcolor{black}{that the deviations from the even Gaussian moments are small and masked by fluctuations whenever the tori diameters are small.} These deviations
led us to a {\it topological} investigation of the distributions $f(q,p,\zeta)$ \textcolor{black}{at different cross sections, that constitutes a much more sensitive indicator of nonergodicity than either the moments or the one-dmensional probabilities associated with them.}  We
carried out visual inspections of zero-$\zeta$ cross sections like those shown in
{\bf Figure 2}. For values of $\nu$ both smaller and larger than the borderline value
$\nu = 0.5$  put forward by Sergi and Ferrario, the sections all revealed well-defined
``holes''.  The ``holes'' that can be seen in the Figure correspond to toroidal solutions
which penetrate the surrounding chaotic sea. 

{\bf Figures 3 and 4}, corresponding to $\nu = +3$ and $\nu = +2.9$ respectively, reveal
triangular regions enclosing nested tori.  These tori are a clear proof of nonergodicity.
The reversibility of the motion equations suggests that changing the signs of
$(\nu,p,\zeta)$ in the initial conditions will simply reverse the trajectories.
Numerical work, using fourth-order, fifth-order, and adaptive Runge-Kutta integrators
bears that expectation out.  The rotation of the triangle seen in the closeups ( with
linear zooms of factors of ten and one hundred ) from longest-side-``up'' to
longest-side-``down'' suggests a singular region in between, which further investigation
locates near $\nu = +2.903521$ .

It appears that the tori shrink to a single periodic orbit at this value of $\nu$
before enlarging again as $\nu$ increases further. The periodic orbit is shown in
{\bf Figure 5}. A zoom into this region by a factor of ten million 
places an upper limit of $3 \times 10^{-8}$ on the size of the thin 
torus that presumably surrounds the periodic orbit. The limiting torus 
has a winding number of (1/3), which means that the orbit rotates through 
an angle $(2 \pi /3)$ the short way around the torus for each time around 
the long way. Since the periodic orbit is a neutrally stable fixed point 
in the Poincar\'e section, it is surrounded by a region that very slowly 
fills in by orbits approaching from the chaotic sea that spend a long 
time in its vicinity, an example of which is evident in Figure 4. Thus 
it appears that at this singular value of $\nu$ the system may be 
ergodic but only after an infinite time.

\begin{figure}
\vspace{1 cm}
\includegraphics[width=4.0in]{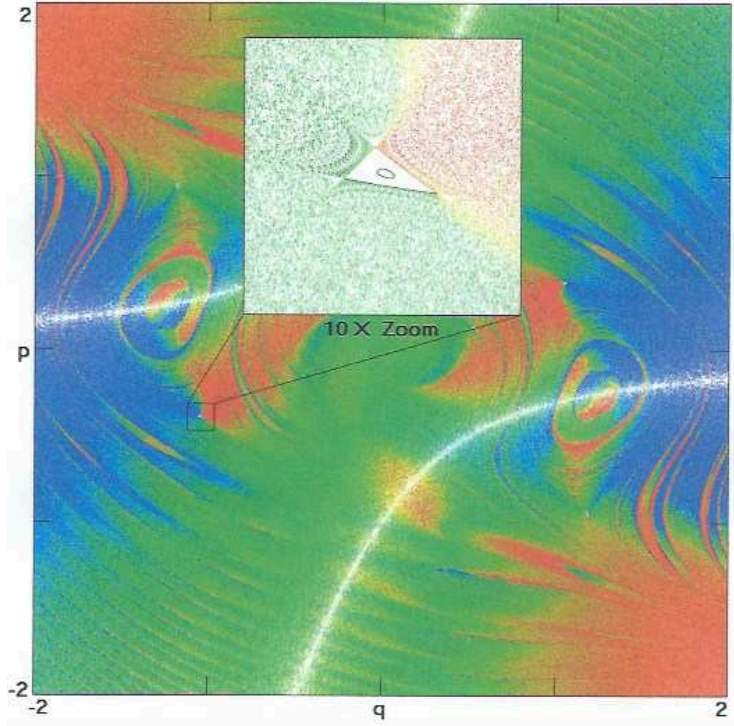}
\caption{
An enlarged version of the $\nu = +3$ case shown in Figure 2 shows four barely-visible
``holes'' in the $\zeta = 0$ cross section.  Magnification of the lower-left hole,
tenfold in both the $q$ and the $p$ directions, reveals a sharp triangular boundary
between nested tori on the inside and chaos on the outside.  Because the location of
the toroidal region is a smooth function of $\nu$ successive approximations track its
center to $\nu = +2.9$ , shown in Figure 4, and finally to $\nu = +2.903521$ , where the
sidelength of the triangle is less than $3 \times 10^{-8}$ .  Color indicates the local value of the
Lyapunov exponent $\lambda_1(t)$ , with red positive and blue negative.
}
\end{figure}

\begin{figure}
\vspace{1 cm}
\includegraphics[width=4.0in]{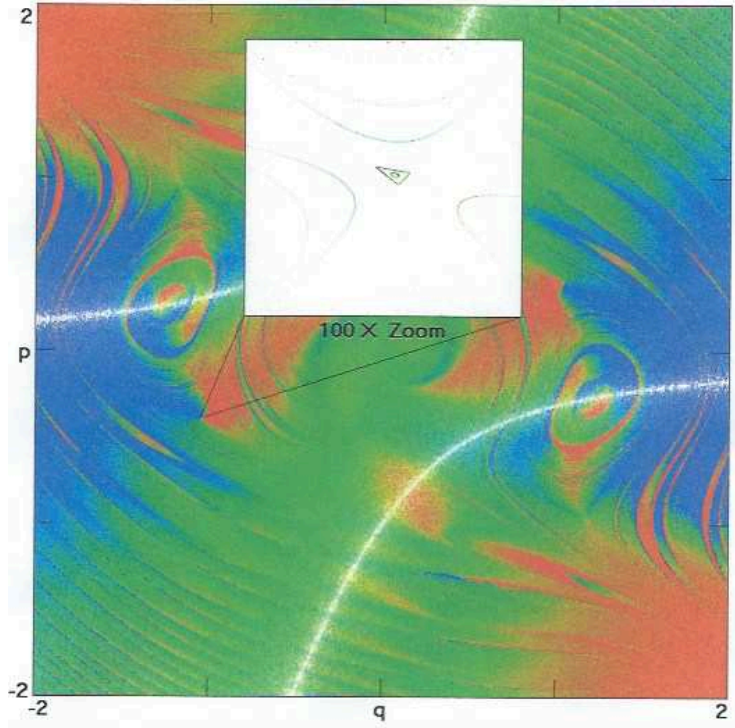}
\caption{
A hundred-fold linear zoom in $q$ and $p$ reveals the toroidal solutions within a triangular
reqion of width of order 0.001 .  The white lines indicate $(q,p)$ tracks of
points moving parallel to the plane.  Red and blue represent positive and negative
local Lyapunov exponents $\lambda_1(t)$ .  The cross section displays the mirror ( or
inversion ) symmetry $(+q,+p) \longleftrightarrow (-q,-p)$ .  $\nu$ is $+2.90$ here.
}
\end{figure}

\begin{figure}
\vspace{1 cm}
\includegraphics[width=4.0in]{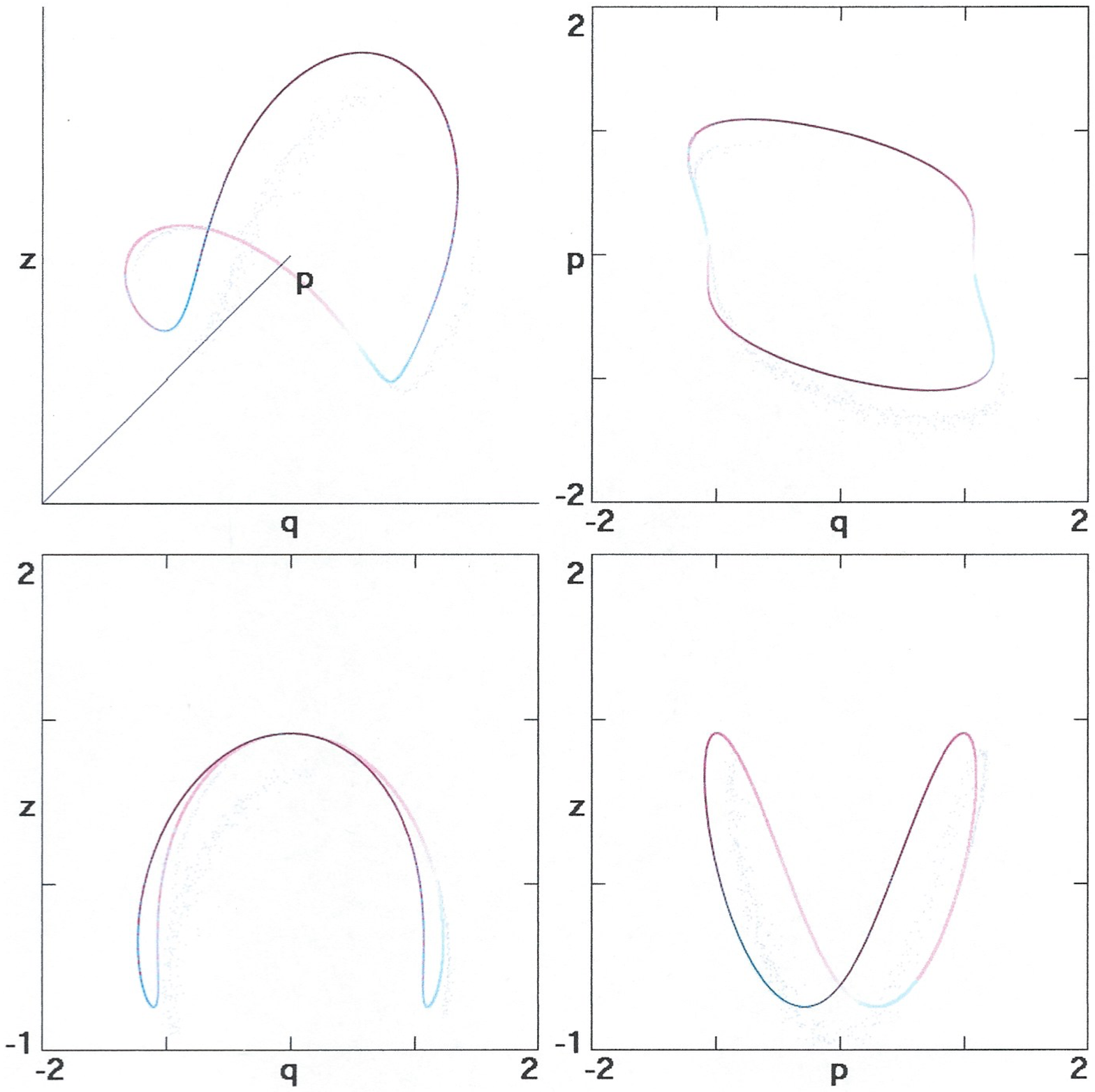}
\caption{
The stable periodic orbit near $\nu = 2.903521$ and its projections into the $(q,p)$,
$(q,\zeta)$, and $(p,\zeta)$ planes.  This orbit has a period of 12.2945528 and
crosses the $\zeta = 0$ plane near $(q, p) = (+1.04099057,-0.395775939)$ .
}
\end{figure}

It seems likely that the mathematical [ as opposed to physical ] form of the oscillator
equations, where $p$ differs from $\dot q$ and where the friction coefficient depends
on $qp$ , was offputting for later investigators so that these 2001 contradictions with
the literature of the 1990s passed either unnoticed or at least undeclared until now.

The systematic explorations carried out by Bauer, Bulgac, Ju, and Kusnezov, for simple
systems including the harmonic oscillator, suggested that {\it quartic} thermostating terms
like $-\zeta p^3$ or $-\zeta^3p$ best promote chaos\cite{b8,b9,b10}.  Accordingly, we modified
the Sergi and Ferrario equations to include a cubic ( rather than linear ) dependence on the
friction coefficient :
$$
\{ \ \dot q = p(1 + \zeta^3 \nu) \ ; \ \dot p = -q -\zeta^3 p \ ; \
\dot \zeta = p^2 - T - qp\nu \ \} \ {\rm [ \ SF' \ ]} \ .
$$
We also followed References 8-10 by considering a quadratic version with $\zeta^3
\rightarrow \zeta|\zeta|$ .  The reader can check to confirm that these equations for the
flow velocity still satisfy the stationary phase-space flow equation ,
$(\partial f/\partial t) = 0 = -\nabla \cdot (fv)$ . Here $v = (\dot q,\dot p,\dot \zeta)$
is the three-dimensional flow velocity in $(qp\zeta)$ space.  The probability densities for
the quadratic and cubic generalized friction coefficients $\zeta$ vary as
$e^{-| \ \zeta \ |^3/3}$ and $e^{-\zeta ^4/4}$ .

\section{Two-Thermostat Harmonic Oscillator Models}

Beginning about 1990 two-thermostat models were developed, mostly based on controlling pairs
of moments\cite{b8,b9,b10,b11,b12}. Applications established the mathematical consistency of
such models with Gibbs' canonical distributions, with barrier-crossing problems, and with
Brownian motion problems. The Ju-Bulgac, Martyna-Klein-Tuckerman, and Hoover-Holian models
all generated the canonical distribution.  The controversial ergodicity of the MKT oscillator
was investigated, and confirmed with particular care, as the 2014 Snook Prize
problem\cite{b13,b14}.  Rather than formulating two controls over the potential or kinetic
energy this MKT ``chain'' model used a second thermostat variable $\xi$ to thermalize the
fluctuations of the first, $\zeta^2$ :
$$
\{ \ \dot q = p \ ; \ \dot p = -q -\zeta p \ ; \
\dot \zeta = p^2 - T - \xi \zeta \ ; \ \dot \xi = \zeta^2 - T \ \} \ {\rm [ \ MKT \ ]} \ 
$$

In 2014 Patra and Bhattacharya discovered that the doubly-thermostated oscillator equations ,
$$
\{ \ \dot q = p - \xi q \ ; \ \dot p = -q -\zeta p \ ; \
\dot \zeta = p^2 - T \ ; \ \dot \xi = q^2 - T \ \} \ {\rm [ \ PB = SE \ ]} \ ,
$$
were not ergodic\cite{b15}.  By coincidence Sergi and Ezra had already found this {\it same}
result in 2010\cite{b16}.  We discovered this by noticing that their Figure 2 looked identical to
Patra and Bhattacharya's Figures 2c and 2d in Reference 15. The key to understanding ergodicity
and its lack in these simple oscillator systems lies in distinguishing two qualitatively
different types of ``holes'' in the cross sections of the flow.  We turn to that next.

\section{Holes in the Singly-Thermostated Cross Sections}

The holes found here in the cross sections are reminiscent of those found recently by
Patra and Bhattacharya\cite{b13}.  They investigated the two unstable fixed points
generated by the four-dimensional MKT oscillator equations.  Evidently the centers of
the largest holes found in the present work typically include fixed cycles of four
repeating points of the mapping from one penetration of the  plane at $\zeta = 0$ to
the next ( three intermediate penetrations separate pairs of point repetitions ) .  The holes
are especially clear for the case $\nu = 2$ in Figure 2.

{\bf Figure 4} illustrates a relatively sensitive case, $\nu = +2.90$ , using the original
Sergi-Ferrario equations with linear friction. Apart from four tiny similar holes in the
section, the chaotic sea outside them has a Gaussian distribution.  This is consistent
with the largest of the long-time-averaged Lyapunov exponents ( as well as with the
complete spectrum of four exponents ) vanishing for all those trajectories which pass
through the holes.

We took the precaution of solving this problem with three different integrators
( fourth-order, fifth-order, and adaptive Runge-Kutta ) and a variety of fixed and
variable timesteps, all in a diligent effort to avoid numerical errors.
For a purely-Hamiltonian harmonic oscillator it is well-known that the fourth-order
method gradually loses energy while the fifth-order method gains.  The good agreement of
all three integrators with one another shows that the nonlinearities of the differential
equations dominate the errors ( on the order of $10^{-17}$ or less at each timestep )
from the finite precision of the simulations.  By simply searching for holes and
evaluating the largest Lyapunov exponents within them, or by evaluating the largest
Lyapunov exponents for millions of randomly-chosen initial conditions it is relatively easy to
separate the chaotic and quasiperiodic regions.

\begin{figure}
\vspace{1 cm}
\includegraphics[width=4.0in,angle=-90.]{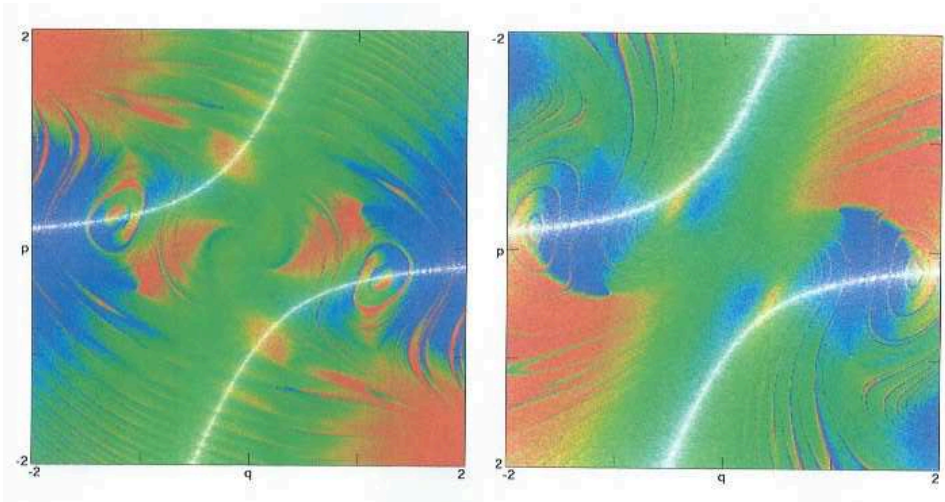}
\caption{
Symmetry breaking at $\nu = \pm 2.9$ .  The time-reversible $(q,\pm p,\pm \zeta,\pm \nu)$
trajectories trace out identical coordinate sequences, but in the opposite time direction.
Even so, differences in the recent past histories of the forward and reversed trajectories
lead to the totally different local values of the largest Lyapunov exponent $\lambda_1(t)$
shown here.  The global forward and backward {\it time averages} of the local exponents
$\lambda_1$ match. The section for $\nu = -2.9$ has been reflected about the coordinate
axis $(p = 0)$ to show that the trajectory penetrations as well as the white nullcline
intersections parallel to the $(q,p,0)$ plane are {\it identical} in the forward and
reversed trajectories.
}
\end{figure}

\section{Summary, Discovery, and Advice}

Three families of singly-thermostated oscillators, with linear, quadratic, and cubic
friction, were formulated to obey Gibbs' canonical distribution for an oscillator.
All provided chaos but none was ergodic. In the linear case it is possible that
there {\it is} a ergodic solution in the vicinity of $\nu = 2.903521$ .  We can
place an upper limit, $\simeq 10^{-16}$ , on the nonergodic measure there. In every
case we examined ( with independent calculations in India, Nevada, and Wisconsin )
the cross sections revealed {\it mixed} solutions, holes containing nested tori
embedded in a chaotic Gaussian sea.  {\bf Figure 5} shows such a stable torus.

  This
situation bears a qualitative resemblance to solutions of the original Nos\'e-Hoover
oscillator.  Visualization and the local values of the largest Lyapunov exponent are
the two most valuable tools for distinguishing the two solution types. To evaluate
the local exponent $\lambda_1(t)$
requires both a ``reference'' trajectory and a nearby ``satellite''.  This increases
the computer time required by only a factor of two.  It is convenient to rescale the
separation between the two similar trajectories ( by displacing the satellite toward
the reference )\cite{b17,b18} at every timestep so as to determine $\lambda_1(t)$ .

These three-dimensional oscillator problems help illuminate features of ergodicity
searches for the four-dimensional flows obtained with two thermostat variables.  Our
work here has shown that the Sergi-Ferrario oscillators are at best seldom ergodic
( assuming only that the hundreds of cases we examined are typical ). 

A particularly fascinating aspect of the fully time-reversible Sergi-Ferrario model
is the symmetry breaking illustrated in {\bf Figure 6}. The forward structure of the flow's
Lyapunov instability is much more complex than the totally different backward structure
despite the fact that any trajectory obeying the SF motion equations can be followed
just as well backward as forward. This ``Arrow of Time'' asymmetry of the largest
Lyapunov exponent $\lambda_1(t)$ for a simple fully time-reversible flow deserves
further study. 

We recommend the challenge of taking up the search for ergodic three-dimensional oscillator
models. In pursuing this elusive goal it seems to us highly desirable to maintain the
conventional relation between the coordinate and the velocity, $\dot q = p$ .  It is also
desirable to resist such physically-artificial accelerations as the $qp$ contribution to the
thermostat variable $\zeta$ .  At their best control variables should utilize transparent
and meaningful origins.  It is certainly possible that with increasing computer power
searches such as those carried out by Sprott\cite{b19} could uncover a host of new
variations of the Sergi-Ferrario or Sergi-Ezra-Patra-Bhattacharya equations which are
simultaneously robust, useful, physically meaningful, and, above all, {\it ergodic}.

In closing, we have recently discovered a particularly promising direction embodying
``weak control'' of the momentum through a choice of the parameters $(\alpha,\beta,\gamma)$
in the set of three moment-based equations of motion :
$$
\{ \ \dot q = +p \ ;
\ \dot p = -q -\zeta[ \ \alpha p + \beta  (p^3/T) +  \gamma (p^5/T^2) \ ] \ ;
$$
$$
\dot \zeta = \alpha[ \ (p^2/T) - 1 \ ] + \beta[ \ (p^4/T^2) - 3(p^2/T) \ ] +
\gamma[ \ (p^6/T^3) - 5(p^4/T^2) \ ] \ \} \ .
$$

Computational searches in $(\alpha,\beta,\gamma)$ space suggest that there {\it are}
regions where the singly-thermostated oscillator samples the entire Gibbs' distribution.
One such combination which appears to be ergodic is $(\alpha,\beta,\gamma) =
(1.50,0.00,-0.50)$ .  We suspect there are many more.  Finding them will conclude a search
set in motion by Shuichi Nos\'e some thirty years ago.

\section{Acknowledgment}
We thank Carol Hoover for her generous help with the figures.


\begin{thebibliography}{99}

\bibitem{b1}  W. G. Hoover, ``Canonical Dynamics: Equilibrium Phase-Space Distributions'', Physical
              Review A {\bf 31}, 1695-1697 (1985).

\bibitem{b2}  S. Nos\'e, ``A Molecular Dynamics Method for Simulations in the Canonical Ensemble'',
              Molecular Physics {\bf 52}, 255-268 (1984).

\bibitem{b3}  S. Nos\'e, ``A Unified Formulation of the Constant Temperature Molecular Dynamics
              Methods'', Journal of Chemical Physics {\bf 81}, 511-519 (1984).

\bibitem{b4}  H. A. Posch, W. G. Hoover, and F. J. Vesely, ``Canonical Dynamics of the Nos\'e
              Oscillator: Stability, Order, and Chaos'', Physical Review A {\bf 33}, 4253-4265
              (1986).

\bibitem{b5}  H. A. Posch and W. G. Hoover, ``Time-Reversible Dissipative Attractors in Three
              and Four Phase-Space Dimensions'', Physical Review E {\bf 55}, 6803-6810 (1997).

\bibitem{b6}  P. K. Patra, J. C. Sprott, W. G. Hoover, and C. G. Hoover, ``Deterministic
              Time-Reversible Thermostats : Chaos, Ergodicity and the Zeroth Law of
              Thermodynamics'', Molecular Physics (in press, 2015).

\bibitem{b7}  A. Sergi and M. Ferrario, ``Non-Hamiltonian Equations of Motion with a Conserved Energy'',
              Physical Review E {\bf 64}, 056125 (2001), Equations 24, 26, and 27.
            
\bibitem{b8}  A. Bulgac and D. Kusnezov, ``Canonical Ensemble Averages from Pseudomicrocanonical
              Dynamics”, Physical Review A {\bf 42}, 5045-5048 (1990).

\bibitem{b9}  D. Kusnezov, A. Bulgac, and W. Bauer, ``Canonical Ensembles from Chaos'', Annals of
              Physics {\bf 204}, 155-185 (1990) and {\bf 214}, 180-218 (1992).

\bibitem{b10} N. Ju and A. Bulgac, ``Finite-Temperature Properties of Sodium Clusters'', Physical
              Review B {\bf 48}, 2721-2732 (1993).

\bibitem{b11} W. G. Hoover and B. L. Holian, ``Kinetic Moments Method for the Canonical Ensemble
              Distribution'', Physics Letters A {\bf 211}, 253-257 (1996).

\bibitem{b12} G. J. Martyna, M. L. Klein, and M. Tuckerman, ``Nos\'e-Hoover Chains: the Canonical
              Ensemble {\it via} Continuous Dynamics'', The Journal of Chemical Physics {\bf 97},
              2635-2643 (1992).

\bibitem{b13} P. K. Patra and B. Bhattacharya, ``Non-Ergodicity of Nos\'e-Hoover Chain Thermostat
              in Computationally Achievable Time'', Physical Review E {\bf 90}, 043304 (2014) =
              ar$\chi$iv : 1407.2353.

\bibitem{b14} W. G. Hoover and C. G. Hoover, ``Ergodicity of the Martyna-Klein-Tuckerman Thermostat
              and the 2014 Snook Prize'', Computational Methods in Science and Technology {\bf 21},
              5-10 (2015) : ar$\chi$iv : 1501.06634.

\bibitem{b15} P. K. Patra and B. Bhattacharya, ``Improving the Ergodic Characteristics of
              Thermostats using Higher Order Temperatures'', ar$\chi$iv : 1411:2194, Figure 2.

\bibitem{b16} A. Sergi and G. S. Ezra, ``Bulgac-Kusnezov-Nos\'e-Hoover Thermostats'', Physical
              Review E {\bf 81}, 036705 (2010), Figure 2.

\bibitem{b17} I. Shimada and T. Nagashima, ``A Numerical Approach to Ergodic Problems of Dissipative
              Dynamical Systems'', Progress of Theoretical Physics {\bf 61}, 1605-1616 (1979).

\bibitem{b18} G. Benettin, L. Galgani, A. Giorgilli, and J. M. Strelcyn, ``Lyapunov Characteristic
              Exponents for Smooth Dynamical Systems and for Hamiltonian Systems; a Method for
              Computing All of Them. Part 1: Theory'', Meccanica {\bf 15}, 9-20 (1980).

\bibitem{b19} J. C. Sprott, ``Some Simple Chaotic Flows'', Physical Review E {\bf 50}, R647-R650 (1994).


\end{thebibliography}
\end{document}